\documentstyle[aps,eqsecnum,preprint,tighten,prd]{revtex}
\begin{document}
\draft
\title{Gravitational radiation from infall into a black hole: \\
       Regularization of the Teukolsky equation}
\author{Eric Poisson$^*$}
\address{Department of Physics, University of Guelph, Guelph,
         Ontario, N1G 2W1, Canada$^\dagger$; \\
         McDonnell Center for the Space Sciences, Department of
         Physics, Washington University, St.~Louis, Missouri,
         63130}
\date{Submitted to Physical Review D, June 27 1996}
\maketitle

\begin{abstract}
The Teukolsky equation has long been known to lead to divergent 
integrals when it is used to calculate the gravitational radiation 
emitted when a test mass falls into a black hole from infinity. Two 
methods have been used in the past to remove those divergent integrals. 
In the first, integrations by parts are carried out, and the infinite 
boundary terms are simply discarded. In the second, the Teukolsky 
equation is transformed into another equation which does not lead to 
divergent integrals. The purpose of this paper is to show that 
there is nothing intrinsically wrong with the Teukolsky equation 
when dealing with non-compact source terms, and that the divergent 
integrals result simply from an incorrect choice of Green's function. 
In this paper, regularization of the Teukolsky equation is carried 
out in an entirely natural way which does not involve modifying the 
equation.
\end{abstract}

\section{Introduction}

In 1971, Davis, Ruffini, Press, and Price \cite{DRPP} computed for the 
first time the amount of energy carried off by gravitational waves 
when a test mass, released from rest at infinity, falls radially 
into a Schwarzschild black hole. By numerically integrating the recently 
derived Zerilli equation \cite{Zerilli} for black-hole perturbations, 
they found an amount $\Delta E = 0.0104 \mu^2/M$, where $\mu$ is the mass
of the infalling particle, and $M$ the mass of the black hole.
(Units are such that $G=c=1$.) 

This calculation was later generalized to many different situations. 
In 1973, Ruffini \cite{Ruffini} considered a particle released from 
infinity with nonvanishing initial velocity. In 1979, Detweiler and 
Szedenits \cite{DetweilerSzedenits} examined infall trajectories 
with nonzero angular momentum. The first calculation involving a 
Kerr black hole was carried out in 1982 by Sasaki and Nakamura 
\cite{SasakiNakamura82}, who considered a particle falling along
the symmetry axis. In 1983, Kojima and Nakamura \cite{KojimaNakamura83} 
examined the case of a particle moving radially within the equatorial 
plane. In 1984, the same authors \cite{KojimaNakamura84} studied infall 
trajectories with nonzero angular momentum. The infall of spinning
test masses was considered for the first time very recently, by Mino, 
Shibata, and Tanaka \cite{MinoShibataTanaka}. The gravitational waves 
emitted by a particle in bound motion around a black hole have also been 
extensively studied \cite{1,2,3,4,5,6,7,8,9,10,11,12,13,14,15,16,17,18}, 
following the pioneering work of Detweiler in 1978 \cite{Detweiler}.

Davis {\it et.~al\/} \cite{DRPP} obtained their classic result by 
integrating the Zerilli equation \cite{Zerilli}, which describes 
in a compact form the (even-parity, or polar) metric perturbations 
of the Schwarzschild black hole. On the other hand, Detweiler and 
Szedenits \cite{DetweilerSzedenits} worked with the Teukolsky 
equation \cite{Teukolsky}, which describes in a compact form the
{\it curvature} perturbations of the Kerr black hole. (In the
limiting case of a nonrotating black hole, and in the absence of
a source for the perturbations, the Teukolsky equation reduces 
to the Bardeen-Press equation \cite{BardeenPress}, which was 
derived earlier.) 

However, when integrating the Teukolsky equation, Detweiler and 
Szedenits encountered divergent integrals, which they regularized 
by integrating by parts and then discarding the infinite boundary 
terms. (The Zerilli formalism does not lead to divergent integrals.) 
This procedure was also adopted by Simone, Poisson, and Will 
\cite{SimonePoissonWill} who, in 1995, reproduced the Davis 
{\it et.~al\/} result using the Teukolsky equation, for the purpose 
of testing the accuracy of post-Newtonian methods. In neither of 
these papers was a justification given for discarding the infinite 
boundary terms, and confidence in the procedure came entirely from 
the comparison with Davis {\it et.~al}. 

The first attempt to remove the divergent integrals from the Teukolsky 
formalism came in 1981 from Tashiro and Ezawa \cite{TashiroEzawa}, who 
employed the clever trick of subtracting from the dependent variable a 
quantity constructed from the source. The resulting equation for the 
new dependent variable leads to well defined integrals. Also in 1981, 
Sasaki and Nakamura \cite{SasakiNakamura81} dealt with the divergent 
integrals in a different way. In the restricted context of 
a Schwarzschild black hole, these authors rewrote the Teukolsky 
equation in a Regge-Wheeler form, with a source term constructed from 
the source of the Teukolsky equation. (The Regge-Wheeler equation 
\cite{ReggeWheeler} describes the odd-parity, or axial, metric 
perturbations of the Schwarzschild black hole.) The Sasaki-Nakamura 
equation, which was later generalized to the case of a Kerr black 
hole \cite{SasakiNakamura82}, does not lead to divergent integrals. 
Subsequent studies of infalling particles 
\cite{KojimaNakamura83,KojimaNakamura84,MinoShibataTanaka} 
were carried out by integrating this equation.

Those methods of regularization give the impression that when 
dealing with unbounded particle trajectories, the Teukolsky 
equation necessarily leads to divergent integrals, and is 
therefore not well posed. The purpose of this paper is to 
show that this impression is based on a misconception. 
Indeed, I wish to show that the Teukolsky equation can be 
regularized in an entirely natural way which does not 
involve modifying the equation. 

The basic issues are easily summarized. 

For perturbations of a Schwarzschild black hole, to which I 
specialize in this paper, and after separation of the variables
(the usual Schwarzschild coordinates are used), the 
(inhomogeneous) Teukolsky equation \cite{Teukolsky} 
takes the form
\begin{equation}
\biggl( \frac{d}{dr}\, p \frac{d}{dr} + p^2 U \biggr) R
= p^2 T.
\label{1.1}
\end{equation}
Here, $R(r)$ is the radial function corresponding to a 
perturbation of frequency $\omega$ and spherical-harmonic 
indices $\ell$ and $m$, $p(r) = (r^2-2Mr)^{-1}$, $U(r)$ 
is the effective potential, whose explicit expression can be
found in Eq.~(\ref{2.15}), and $T(r)$ is the source term, 
which is constructed from the energy-momentum tensor 
of the infalling mass. 

The inhomogeneous Teukolsky equation is solved with
the physical requirement that gravitational waves must be
purely ingoing at the black-hole horizon, and purely outgoing
at infinity; this is equivalent to a no-incoming-radiation 
initial condition. Mathematically, this translates into
two statements. First, that $R(r) \propto R^H(r)$ near $r=2M$. 
Here, $R^H(r)$ is a solution to the homogeneous equation, normalized 
such that $R^H(r\to 2M) \sim (1-2M/r)^2 \exp(-i\omega r^*)$, where
$r^* = r + 2M\ln(r/2M-1)$. Second, that $R(r) \propto R^\infty(r)$
near $r=\infty$, where $R^\infty(r)$ is also a solution to the
homogeneous equation, normalized such that $R^\infty(r\to\infty)
\sim (i\omega r)^3 \exp(i\omega r^*)$. 

The most convenient way of integrating Eq.~(\ref{1.1}) is by means 
of a Green's function, which should be chosen so as to incorporate
the specified boundary conditions. The standard theory \cite{Arfken} 
{\it suggests} that the appropriate solution is 
\begin{equation}
R(r) = \frac{R^\infty(r)}{W_{\rm T}}\, 
\int_{2M}^r p^2(r') T(r') R^H(r')\, dr'
+ \frac{R^H(r)}{W_{\rm T}}\, 
\int_r^\infty p^2(r') T(r') R^\infty(r')\, dr',
\label{1.2}
\end{equation}
where $W_{\rm T}$ is a constant, equal to the conserved Wronskian 
of $R^H(r)$ and $R^\infty(r)$. The misconception is precisely that 
Eq.~(\ref{1.2}) must be the desired solution. In fact it is not: 
In order for the standard theory of Green's functions to guarantee 
that $R(r)$ as given by Eq.~(\ref{1.2}) satisfies the specified 
boundary conditions, {\it the integrals must converge}. 
This is not the case here. Because $R^H(r)$ and $R^\infty(r)$ 
both have a component growing as $r^3$ when $r\to\infty$, and 
because $p^2(r) T(r)$ falls off only as $r^{-3/2}$, those integrals
diverge when $r\to \infty$. The claim that Eq.~(\ref{1.2})
enforces the specified boundary conditions is 
therefore unjustified, and in fact, is wrong. 

The method adopted in this paper for regularizing the Teukolsky
equation goes as follows. For simplicity, I consider the
specific case of a particle falling radially into a Schwarzschild
black hole, having been released from rest at infinity. Generalization
to other situations should be straightforward.

Instead of the ill-defined particular solution (\ref{1.2}), 
the starting point of this analysis is the most general solution
to Eq.~(\ref{1.1}), which is written as 
\begin{equation}
R(r) =  \frac{R^\infty(r)}{W_{\rm T}}\, \biggl[ A + 
\int_{a}^r p^2(r') T(r') R^H(r')\, dr' \biggr]
 + \frac{R^H(r)}{W_{\rm T}}\, \biggl[ B + 
\int_r^b p^2(r') T(r') R^\infty(r')\, dr' \biggr],
\label{1.3}
\end{equation}
where $a$, $b$, $A$, and $B$ are constants. This represents 
the gravitational radiation generated by the infalling particle, 
plus the free radiation that was initially present in the spacetime. 
Boundary conditions must be imposed to eliminate this free 
component.

Now, because the integrals are ill defined when $r\to\infty$,
these conditions cannot be imposed immediately. Instead, 
integrations by parts are carried out, which are specifically 
designed to  make the integrals well behaved. The boundary 
terms at $r'=a$ and $r'=b$ are then absorbed into the constants 
$A$ and $B$, while the boundary terms at $r'=r$ are carefully kept. 
At this stage one is still dealing with the most general 
solution to the Teukolsky equation, and this is a perfectly
valid starting point for the discussion of boundary conditions.
Since the integrals are now well behaved, there is no obstacle 
in setting $a=2M$ and $b=\infty$. The new constants $A$ and $B$ 
are then chosen so that the solution satisfies the specified 
boundary conditions. By this procedure, regularization of 
the Teukolsky equation involves no unjustified manipulations
nor modifications to the equation, and is entirely natural. 

The idea described in this paper appears to be completely trivial,
and it is indeed ironic that such a solution to the problem of 
divergent integrals did not come forth much sooner. However, 
the work required to carry out the procedure is not, in itself, 
entirely trivial. The rest of the paper is devoted to a detailed
presentation. It is organized as follows:

For the purpose of integrating by parts, it is convenient to
write the Teukolsky functions $R^H(r)$ and $R^\infty(r)$ in
terms of the related, and better behaved, Regge-Wheeler 
functions $X^H(r)$ and $X^\infty(r)$. These functions, and 
the transformations relating them, are described in detail 
in Sec.~II, which establishes many results used later on. 

Regularization of the inhomogeneous Teukolsky equation is 
carried out in Sec.~III for the specific case of a particle 
falling radially into a Schwarzschild black hole. Generalization
to other cases should proceed along similar lines, but I shall 
not pursue this here.

The main results of this paper are summarized in Sec.~IV.

{\it Notation.} The following symbols appear frequently 
throughout the paper: $f=1-2M/r$, $r^*=r+2M\ln(r/2M-1)$, 
$d/dr^* = f d/dr$, $Q=iM\omega$, $z=(i\omega r)^{-1}$,
and $x=(r/2M)^{1/2}$. 

\section{Regge-Wheeler, Teukolsky, and Chandrasekhar}

This section is devoted to the derivation of various
results which will be used in the following section.
Specifically, Sec.~II A contains a discussion of the
Regge-Wheeler equation \cite{ReggeWheeler}, and a study 
of the asymptotic behavior of its solutions near $r=2M$ 
and near $r=\infty$. Section II B does the same for the 
homogeneous Teukolsky equation \cite{Teukolsky}. The 
Chandrasekhar transformation \cite{Chandrasekhar}, which 
relates a solution to the Regge-Wheeler equation to a solution 
of the homogeneous Teukolsky equation, is the topic of 
Sec.~II C. Finally, the functions $X^H(r)$, $X^\infty(r)$,
$R^H(r)$, $R^\infty(r)$, and the relations between them, 
are introduced in Sec.~II D.

\subsection{Regge-Wheeler equation}

The Regge-Wheeler equation \cite{ReggeWheeler} compactly describes 
a metric perturbation  of frequency $\omega$ and spherical-harmonic 
indices $\ell$ and $m$. It reads
\begin{equation}
\Biggl\{ \frac{d^2}{dr^{*2}} + \omega^2 - f
\biggl[ \frac{\ell(\ell+1)}{r^2} - \frac{6M}{r^3} \biggr]
\Biggr\} X(r) = 0,
\label{2.1}
\end{equation}
where $f=1-2M/r$ and $d/dr^* = fd/dr$, so that 
\begin{equation}
r^* = r + 2M\ln(r/2M-1).
\label{2.2}
\end{equation}
For two linearly independent solutions $X_1(r)$ and $X_2(r)$, the 
conserved Wronskian is given by
\begin{equation}
W_{\rm RW} (X_1,X_2) = f \bigl( X_1 X_2' - X_2 X_1' \bigr),
\label{2.3}
\end{equation}
where a prime denotes differentiation with respect to $r$.
Because the Regge-Wheeler equation is real, if $X_1(r)$ is
a solution, then $X_2(r) = \bar{X}_1(r)$ is also a
solution, linearly independent from the first;
the overbar denotes complex conjugation.

\subsubsection{Asymptotic behavior near $r=2M$}

It can be seen from Eq.~(\ref{2.1}) that near $r=2M$, 
the Regge-Wheeler function must behave as 
$\exp(-i\omega r^*)$ or its complex conjugate.
To obtain more information, let
\begin{equation}
X(r) = Y(f)\, e^{-i\omega r^*}.
\label{2.4}
\end{equation}
Then a short calculation shows that the new function 
$Y(f)$ must satisfy
\begin{equation}
(1-f)^2 f\, Y'' + \bigl[ (1-f)(1-3f) - 4Q \bigr]\, Y'
 - \bigl[\ell(\ell+1)-3(1-f) \bigr]\, Y = 0,
\label{2.5}
\end{equation}
where $Q \equiv iM\omega$ and a prime denotes differentiation
with respect to $f$. Equation (\ref{2.5}) can be integrated by 
by writing
\begin{equation}
Y(f) = 1 + \sum_{n=1}^{\infty} a_n f^n,
\label{2.6}
\end{equation}
where the normalization was chosen arbitrarily. Substituting
this into Eq.~(\ref{2.5}) and setting each term of the
resulting series to zero gives
\begin{eqnarray}
a_1 &=& \frac{ \ell(\ell+1)-3 }{1-4Q}, 
\nonumber \\
& & \label{2.7} \\
a_2 &=& \frac{ (\ell-1)\ell(\ell+1)(\ell+2)-12Q }{4(1-2Q)(1-4Q)}.
\nonumber
\end{eqnarray}
The higher-order coefficients can be obtained from the
recurrence relation
\begin{equation}
n(n-4Q)\, a_n = \bigl[ 2(n-1)n + \ell(\ell+1) -3 \bigr]\, a_{n-1}
- (n-3)(n+1)\, a_{n-2},
\label{2.8}
\end{equation}
which holds for $n\geq 3$. 

\subsubsection{Asymptotic behavior near $r=\infty$}

Near $r=\infty$, the Regge-Wheeler function must also 
behave as $\exp(-i\omega r^*)$ or its complex conjugate.
As before, let
\begin{equation}
X(r) = Y(z)\, e^{-i\omega r^*},
\label{2.9}
\end{equation}
where $z=(i\omega r)^{-1}$. This leads to the
differential equation
\begin{equation}
z^2 (1-2Q z)\, Y'' + 2(1+z-3Qz^2)\, Y' 
 - \bigl[ \ell(\ell+1) - 6Qz \bigr]\, Y = 0 ,
\label{2.10}
\end{equation}
where a prime denotes differentiation with respect to $z$.
Equation (\ref{2.10}) is solved by writing
\begin{equation}
Y(z) = 1 + \sum_{n=1}^{\infty} b_n z^n,
\label{2.11}
\end{equation}
where once again the normalization was chosen arbitrarily.  
This gives 
\begin{eqnarray}
b_1 &=& {\textstyle \frac{1}{2}} \ell(\ell+1), 
\nonumber \\
& & \label{2.12} \\
b_2 &=& {\textstyle \frac{1}{8}} \bigl[ (\ell-1)
\ell (\ell+1) (\ell+2) -12Q \bigr].
\nonumber
\end{eqnarray}
All other coefficients can be obtained from the
recurrence relation
\begin{equation}
2n\, b_n = (\ell-n+1)(\ell+n)\, b_{n-1} 
+ 2(n-3)(n+1)Q\, b_{n-2},
\label{2.13}
\end{equation}
which holds for $n\geq 3$. 

\subsection{Homogeneous Teukolsky equation}

The homogeneous Teukolsky equation \cite{Teukolsky} compactly
describes a curvature perturbation of frequency $\omega$ and 
spherical-harmonic indices $\ell$ and $m$, and is given by
\begin{equation}
\biggl[ \frac{d}{dr}\, p(r) \frac{d}{dr} +
p^2(r) U(r) \biggr] R(r) = 0,
\label{2.14}
\end{equation}
where
\begin{eqnarray}
p(r) &=& \frac{1}{r^2 f}, \nonumber \\
& & \label{2.15} \\
U(r) &=& f^{-1} \bigl[ (\omega r)^2 - 4 i \omega (r-3M) \bigr]
- (\ell-1)(\ell+2). \nonumber
\end{eqnarray}
If $R_1(r)$ and $R_2(r)$ are two linearly independent solutions,
then
\begin{equation}
W_{\rm T}(R_1,R_2) = p \bigl( R_1 R_2' - R_2 R_1' \bigr), 
\label{2.16}
\end{equation}
where a prime denotes differentiation with respect to $r$,
is the conserved Wronskian. It should noted that the Teukolsky
equation is complex; the complex conjugate of a solution is
therefore not a solution. For this reason, an ingoing mode 
of the Teukolsky equation, which is proportional to 
$\exp(-i \omega r^*)$, must be distinguished from an
outgoing mode, proportional to $\exp(i \omega r^*)$.

\widetext
\subsubsection{Ingoing mode: Asymptotic behavior near $r=2M$}

It is easy to see from Eq.~(\ref{2.14}) that near the horizon,
an ingoing mode must behave as $f^2 \exp(-i \omega r^*)$. This
motivates the substitution
\begin{equation}
R(r) = f^2 S^{\rm in}(f) e^{-i\omega r^*},
\label{2.17}
\end{equation}
which implies 
\begin{eqnarray}
0 &=& (1-f)^3 f\, S^{{\rm in}\prime\prime} + 
(1-f)(7f^2 - 10f + 3 - 4Q)\, S^{{\rm in}\prime} 
\nonumber \\ \mbox{} & &
- \bigl[ 8f^2 - (\ell^2+\ell+14)f + 
\ell^2+\ell+6+4Q \bigr]\, S^{\rm in}.
\label{2.18}
\end{eqnarray}
Here, a prime denotes differentiation with respect to $f$. A
solution can be obtained by writing
\begin{equation}
S^{\rm in}(f) = 1 + \sum_{n=1}^{\infty} p^{\rm in}_n f^n,
\label{2.19}
\end{equation}
which gives
\begin{eqnarray}
p^{\rm in}_1 &=& \frac{\ell(\ell+1)+6+4Q}{3-4Q}, \nonumber \\
& & \label{2.20} \\
p^{\rm in}_2 &=& \frac{\bigl[ \ell(\ell+1) + 4\bigr]
\bigl[\ell(\ell+1)+18 \bigr] + 4 \bigl[ 2\ell(\ell+1)+33 
\bigr] Q}{8(1-Q)(3-4Q)}. \nonumber
\end{eqnarray}
All other coefficients can be obtained from the recurrence
relation
\begin{eqnarray}
n(n+2-4Q)\, p^{\rm in}_n &=& \bigl[ n(3n+4) + \ell(\ell+1) -1
- 4(n-2)Q \bigr] \, p^{\rm in}_{n-1} 
\nonumber \\ & & \mbox{}
- \bigl[ n(3n+2) + (\ell-1)(\ell+2) \bigr]\, p^{\rm in}_{n-2} 
+ (n^2-1)\, p^{\rm in}_{n-3},
\label{2.21}
\end{eqnarray}
which is valid for $n\geq 3$ (with $p^{\rm in}_0 \equiv 1$).

\subsubsection{Outgoing mode: Asymptotic behavior near $r=2M$}

Equation (\ref{2.14}) implies that near the horizon,
an outgoing mode must behave as $\exp(-i \omega r^*)$. This
leads the substitution
\begin{equation}
R(r) = S^{\rm out}(f) e^{i\omega r^*},
\label{2.22}
\end{equation}
which implies
\begin{eqnarray}
0 &=& (1-f)^3 f\, S^{{\rm out}\prime\prime} + 
(1-f)(3f^2 - 2f -1 + 4Q)\, S^{{\rm out}\prime} 
\nonumber \\ & & \mbox{}
- \bigl[ (\ell-1)(\ell+2)(1-f) +  12Q \bigr]\, S^{\rm out} = 0,
\label{2.23}
\end{eqnarray}
where a prime denotes differentiation with respect to $f$.
This is integrated by writing
\begin{equation}
S^{\rm out}(f) = 1 + \sum_{n=1}^{\infty} p^{\rm out}_n f^n,
\label{2.24}
\end{equation}
which gives
\begin{eqnarray}
p^{\rm out}_1 &=&- \frac{(\ell-1)(\ell+2)+12Q}{1-4Q}, 
\nonumber \\
& & \label{2.25} \\
p^{\rm out}_2 &=& -\frac{(\ell-1)\ell(\ell+1)(\ell+2) 
+ 12\bigl[ 2\ell(\ell+1) -3 \bigr] Q + 
192Q^2}{8Q(1-4Q)}. \nonumber
\end{eqnarray}
All other coefficients can be determined with the recurrence
relation
\begin{eqnarray}
n(n-2+4Q)\, p^{\rm out}_n &=& \bigl[ (n-1)(3n-5) + 
(\ell-1)(\ell+2) + 4(n-2)Q \bigr]\, p^{\rm out}_{n-1} 
\nonumber \\ & & \mbox{} \hspace{-6pt} - 
\bigl[ (n-2)(3n-4) + (\ell-1)(\ell+2) \bigr]\, p^{\rm out}_{n-2} 
+ (n-3)(n-1)\, p^{\rm out}_{n-3},
\label{2.26}
\end{eqnarray}
which is valid for $n\geq 3$ (with $p^{\rm out}_0 \equiv 1$).

\subsubsection{Ingoing mode: Asymptotic behavior near $r=\infty$}

Near infinity, an ingoing mode of the Teukolsky equation must
behave as $r^{-1} \exp(-i\omega r^*)$, which suggests the
substitution
\begin{equation}
R(r) = z S^{\rm in}(z) e^{-i\omega r^*},
\label{2.27}
\end{equation}
where $z= (i\omega r)^{-1}$. The homogeneous Teukolsky
equation then implies
\begin{eqnarray}
0 &=& (1-2Qz)^2 z^2\, S^{{\rm in}\prime\prime} + 
2(1-2Qz)(1+3z-5Qz^2)\, S^{{\rm in}\prime}
\nonumber \\ & & \mbox{}
+ \bigl[ -(\ell-2)(\ell+3) + 8Q + 2(\ell^2+\ell-9)Qz + 
12 Q^2 z^2\bigr]\, S^{\rm in} = 0,
\label{2.28}
\end{eqnarray}
where a prime denotes differentiation with respect to $z$.
Once again the solution is written as a series,
\begin{equation}
S^{\rm in}(z) = 1 + \sum_{n=1}^\infty q^{\rm in}_n z^n,
\label{2.29}
\end{equation}
and substitution yields
\begin{eqnarray}
q^{\rm in}_1 &=& {\textstyle \frac{1}{2}} \bigl[ 
(\ell-2)(\ell+3) - 8Q \bigr], \nonumber \\
& & \label{2.30} \\
q^{\rm in}_2 &=& {\textstyle \frac{1}{8}} \Bigl\{
(\ell-3)(\ell-2)(\ell+3)(\ell+4) - 4 \bigl[ 4\ell(\ell+1) 
-39 \bigl]Q + 32Q^2 \Bigr\}. \nonumber
\end{eqnarray}
The recurrence relation
\begin{eqnarray}
2nq^{\rm in}_n &=& \bigl[ -n(n+3)+(\ell-1)(\ell+2) + 
4(n-3)Q \bigr]\, q^{\rm in}_{n-1} 
\nonumber \\ & & \mbox{} + 
2\bigl[ n(2n+1)-\ell(\ell+1)-1\bigr]Q\, q^{\rm in}_{n-2}
- 4n(n+2) Q^2\, q^{\rm in}_{n-3},
\label{2.31}
\end{eqnarray}
valid for $n\geq 3$ (with $q^{\rm in}_0 \equiv 1$), 
gives the remaining coefficients. 

\subsubsection{Outgoing mode: Asymptotic behavior near $r=\infty$}

Finally, an outgoing mode of the Teukolsky equation behaves as 
$r^{3} \exp(i\omega r^*)$ near infinity. This leads to
\begin{equation}
R(r) = z^{-3} S^{\rm out}(z) e^{i\omega r^*}.
\label{2.32}
\end{equation}
Equation (\ref{2.14}) then implies
\begin{equation}
(1-2Qz) z^2\, S^{{\rm out}\prime\prime} 
- (2+2z-6Qz^2)\, S^{{\rm out}\prime}
- \bigl[(\ell-1)(\ell+2) + 6Qz \bigr]\, S^{\rm out} = 0,
\label{2.33}
\end{equation}
where a prime denotes differentiation with respect to $z$.
This is integrated by substituting
\begin{equation}
S^{\rm out}(z) = 1 + \sum_{n=1}^\infty q^{\rm out}_n z^n
\label{2.34}
\end{equation}
into the differential equation, which gives
\begin{eqnarray}
q^{\rm out}_1 &=& -{\textstyle \frac{1}{2}} 
(\ell-1)(\ell+2), \nonumber \\
& & \label{2.35} \\
q^{\rm out}_2 &=& {\textstyle \frac{1}{8}} \bigl[
(\ell-1)\ell(\ell+1)(\ell+2) - 12 Q \bigr]. \nonumber
\end{eqnarray}
The other coefficients are generated by the 
recurrence relation
\begin{equation}
2n\, q^{\rm out}_n = \bigl[ (n-4)(n-1) - (\ell-1)(\ell+2) 
\bigr]\, q^{\rm out}_{n-1} - 
2 (n-3)(n-5)Q\, q^{\rm out}_{n-2},
\label{2.36}
\end{equation}
which is valid for $n\geq 3$.

\narrowtext
\subsection{Chandrasekhar transformation}

In 1975, Chandrasekhar \cite{Chandrasekhar} proved the following 
theorem: If $X(r)$ is a solution to the Regge-Wheeler equation 
(\ref{2.1}), then there exists a linear differential operator 
$\cal C$ such that $R(r) = {\cal C} X(r)$ is a solution to the 
homogeneous Teukolsky equation (\ref{2.14}). 
The {\it Chandrasekhar transformation} 
is given by ${\cal C} \propto r^2 f {\cal L} f^{-1} {\cal L}
r$, where ${\cal L} = fd/dr + i\omega$. Since $X(r)$ satisfies 
a second-order differential equation, ${\cal C}$ can also be 
written in first-order form as \cite{15}
\begin{equation}
{\cal C} = (i\omega r)\Bigl\{ 2(1-3M/r+i\omega r)r{\cal L} 
+ f \bigl[\ell(\ell+1)-6M/r\bigr] \Bigr\}.
\label{2.37}
\end{equation}
The constant of proportionality was chosen arbitrarily. 

Using the results derived in Sec.~II A, simple 
manipulations are required to prove the following 
statements. First, concerning asymptotic relations 
near $r=2M$:
\begin{eqnarray}
& & \mbox{If $X \sim e^{-i \omega r^*}$, 
then ${\cal C} X \sim 4 Q a_2 \, f^2 e^{-i \omega r^*}$.} 
\label{2.38} \\
& & \mbox{If $X \sim e^{i \omega r^*}$, 
then ${\cal C} X \sim -8Q^2(1-4Q)\, e^{i \omega r^*}$.} 
\label{2.39}
\end{eqnarray}
Next, concerning asymptotic relations near 
$r=\infty$:
\begin{eqnarray}
& & \mbox{If $X \sim e^{-i \omega r^*}$, 
then ${\cal C} X \sim 2b_2\, (i\omega r)^{-1} 
e^{-i \omega r^*}$.}
\label{2.40} \\
& & \mbox{If $X \sim e^{i \omega r^*}$, 
then ${\cal C} X \sim 4\, (i\omega r)^{3} e^{i \omega r^*}$.} 
\label{2.41}
\end{eqnarray}
The constants $a_2$ and $b_2$ are given by Eqs.~(\ref{2.7}) and
(\ref{2.12}), respectively. As they must, the asymptotic relations
found here for ${\cal C} X(r)$ agree with the relations derived 
for $R(r)$ in subsection B. 

\subsection{Linearly independent solutions}

\subsubsection{Asymptotic relations}

Of all the solutions to the Regge-Wheeler equation, two
are preferred. The first describes gravitational waves
which are purely ingoing at the black-hole horizon, 
and is denoted $X^H(r)$. The other describes waves which 
are purely outgoing at infinity, and is denoted 
$X^\infty(r)$. These solutions satisfy the asymptotic 
relations
\begin{equation}
X^H(r) \sim \left\{
\begin{array}{ll}
e^{-i\omega r^*} & \ r \to 2M \\
{\cal A}^{\rm in}\, e^{-i\omega r^*} + 
{\cal A}^{\rm out}\, e^{i\omega r^*} & \ r\to \infty
\end{array} \right. ,
\label{2.42}
\end{equation}
where ${\cal A}^{\rm in}$ and ${\cal A}^{\rm out}$ are
constants, and
\begin{equation}
X^\infty(r) \sim \left\{
\begin{array}{ll}
e^{i\omega r^*} & \ r \to \infty \\
{\cal B}^{\rm in}\, e^{-i\omega r^*} + 
{\cal B}^{\rm out}\, e^{i\omega r^*} & \ r\to 2M
\end{array} \right.,
\label{2.43}
\end{equation}
where ${\cal B}^{\rm in}$ and ${\cal B}^{\rm out}$ are
also constants. These solutions are linearly independent. 
Evaluation of their Wronskian in the limit $r\to\infty$ 
indeed reveals that
\begin{equation}
W_{\rm RG}(X^H,X^\infty) = 2i\omega {\cal A}^{\rm in},
\label{2.44}
\end{equation}
where $W_{\rm RW}$ was defined in Eq.~(\ref{2.3}).

Acting with the Chandrasekhar transformation,
\begin{eqnarray}
R^H(r) &=& \chi^H\, {\cal C} X^H(r), 
\nonumber \\
& & \label{2.45} \\
R^\infty(r) &=& \chi^\infty\, {\cal C} X^\infty(r),
\nonumber
\end{eqnarray}
where $\chi^H$ and $\chi^\infty$ are normalization constants,
returns solutions to the homogeneous Teukolsky equation 
possessing the same physical interpretation. 
These are normalized so that
\begin{equation}
R^H(r) \sim \left\{
\begin{array}{ll}
f^2 e^{-i\omega r^*} & \ r \to 2M \\
{\cal Q}^{\rm in}\, (i\omega r)^{-1} e^{-i\omega r^*} 
+ {\cal Q}^{\rm out}\, (i\omega r)^3 e^{i\omega r^*} 
& \ r\to \infty
\end{array} \right. ,
\label{2.46}
\end{equation}
where ${\cal Q}^{\rm in}$ and ${\cal Q}^{\rm out}$ are
constants, and
\begin{equation}
R^\infty(r) \sim \left\{
\begin{array}{ll}
(i\omega r)^3 e^{i\omega r^*} & \ r \to \infty \\
{\cal P}^{\rm in}\, f^2 e^{-i\omega r^*} 
 + {\cal P}^{\rm out}\, e^{i\omega r^*} & \ r\to 2M
\end{array} \right.,
\label{2.47}
\end{equation}
where ${\cal P}^{\rm in}$ and ${\cal P}^{\rm out}$ are
also constants. It follows from Eqs.~(\ref{2.38}) and 
(\ref{2.41}) that this normalization is obtained by 
choosing
\begin{eqnarray}
\chi^H &=& \frac{1}{4Qa_2} = 
\frac{(1-2Q)(1-4Q)}{Q\bigl[ (\ell-1)\ell(\ell+1)(\ell+2)-12Q\bigr]}, 
\nonumber \\
& & \label{2.48} \\
\chi^\infty &=& \frac{1}{4}. \nonumber
\end{eqnarray}
These solutions are also linearly independent, and
\begin{equation}
W_{\rm T}(R^H,R^\infty) = -2 i \omega^3 {\cal Q}^{\rm in},
\label{2.49}
\end{equation}
where $W_{\rm T}$ was defined in Eq.~(\ref{2.16}). 

\subsubsection{Relations among constants}

The constants ${\cal A}^{\rm in,out}$, ${\cal B}^{\rm in,out}$,
${\cal Q}^{\rm in,out}$, and ${\cal P}^{\rm in,out}$ are not
all independent. Various relations among them are easily 
derived.

In Eq.~(\ref{2.44}), the Wronskian $W_{\rm RW}(X^H,X^\infty)$ 
was evaluated near $r\to\infty$. It can also be evaluated near 
$r=2M$. Since the two values must agree, we have
\begin{equation}
{\cal B}^{\rm out} = {\cal A}^{\rm in}.
\label{2.50}
\end{equation}
Similarly, constancy of $W_{\rm RW}(X^H,\bar{X}^\infty)$
implies
\begin{equation}
{\cal B}^{\rm in} = - \bar{{\cal A}}^{\rm out},
\label{2.51}
\end{equation}
while constancy of $W_{\rm RW}(X^H,\bar{X}^H)$ gives
\begin{equation}
|{\cal A}^{\rm in}|^2 - |{\cal A}^{\rm out}|^2 = 1,
\label{2.52}
\end{equation}
which expresses global conservation of energy. 

Additional relations are a consequence of the Chandrasekhar
transformation. Combining Eqs.~(\ref{2.40})--(\ref{2.42}),
(\ref{2.45}), (\ref{2.46}), and (\ref{2.48}) reveals that
\begin{eqnarray}
{\cal Q}^{\rm in} &=& \frac{b_2}{2Q a_2}\, {\cal A}^{\rm in}
= \frac{(1-2Q)(1-4Q)}{4Q}\, {\cal A}^{\rm in},
\nonumber \\ 
& & \label{2.53} \\
{\cal Q}^{\rm out} &=& \frac{1}{Q a_2}\, {\cal A}^{\rm out}.
\nonumber
\end{eqnarray}
Similarly, combining Eqs.~(\ref{2.38}), (\ref{2.39}),
(\ref{2.43}), (\ref{2.45}), (\ref{2.47}), and (\ref{2.48})
gives
\begin{eqnarray}
{\cal P}^{\rm in} &=& Q a_2\, {\cal B}^{\rm in},
\nonumber \\ 
& & \label{2.54} \\
{\cal P}^{\rm out} &=& -2Q^2(1-4Q)\, {\cal B}^{\rm out}.
\nonumber
\end{eqnarray}
Finally, combining Eqs.~(\ref{2.50}), (\ref{2.51}), 
(\ref{2.53}), and (\ref{2.54}) yields
\begin{eqnarray}
{\cal P}^{\rm out} &=& -\frac{8Q^3}{1-2Q}\, {\cal Q}^{\rm in},
\nonumber \\ 
& & \label{2.55} \\
{\cal P}^{\rm in} &=& Q^2 |a_2|^2\, 
\bar{{\cal Q}}^{\rm out}.
\nonumber
\end{eqnarray}
This last equation does not follow easily from Wronskian
relations.

\section{Regularization of the Teukolsky equation}

I now proceed with the regularization of the Teukolsky equation.
The source function $T(r)$ is constructed in Sec.~III A for
the specific case of a particle of mass $\mu$ released from 
rest at infinity and falling with zero angular momentum into 
a Schwarzschild black hole of mass $M$. The general solution to
the inhomogeneous Teukolsky equation is also displayed here. In
Sec.~III B the regularization procedure is carried out. Then the
behavior of the regularized solution is examined near $r=2M$
in Sec.~III C, and near $r=\infty$ in Sec.~III D.

\subsection{Source term and general solution}

The source function is easily constructed by following the 
steps spelled out in Poisson and Sasaki \cite{15}. 
When the motion is purely radial, it is given by
\begin{equation}
T(r) = 2\sqrt{(\ell-1)\ell(\ell+1)(\ell+2)}\, r^4
\int dt' d\Omega'\, T_{\alpha\beta} n^\alpha n^\beta\, 
\bar{Y}_{\ell m}(\theta',\phi')\, e^{i\omega t'},
\label{3.1}
\end{equation}
where $d\Omega' = d\cos\theta' d\phi'$, $n^\alpha = 
\frac{1}{2} (1,f,0,0)$ is a null vector pointing 
outward, $Y_{\ell m}$ are the usual spherical harmonics, 
and $T^{\alpha \beta}$ is the particle's 
energy-momentum tensor,
\begin{equation}
T^{\alpha\beta}(x') = \mu \int d\tau\, u^\alpha
u^\beta\, \delta\bigl[x'-x(\tau)\bigr].
\label{3.2}
\end{equation}
Here, $x'$ represents an event in spacetime, labeled
by the Schwarzschild coordinates $(t',r',\theta',\phi')$,
and $x(\tau)$ represents the particle's world line, with
four-velocity $u^\alpha=dx^\alpha/d\tau$, where $\tau$ is 
proper time. In Eq.~(\ref{3.2}), the $\delta$-function is
normalized so that $\int \delta(x) \sqrt{-g}\, d^4 x = 
1$, where $g$ is the determinant of the metric. 

The geodesic equations for radial motion reduce to
$\theta=\phi=0$ and
\begin{equation}
\frac{dt}{dr} = - \frac{1}{f} 
\biggl(\frac{r}{2M}\biggr)^{1/2},
\label{3.3}
\end{equation}
which integrates to
\begin{equation}
t(r) = -2M \biggl( {\textstyle \frac{2}{3}} x^3 + 2x +
\ln \frac{x-1}{x+1} \biggr),
\label{3.4}
\end{equation}
where $x\equiv (r/2M)^{1/2}$. The four-velocity has
non-vanishing components $u^t = 1/f$ and $u^r = - 1/x$.

To obtain the source, Eq.~(\ref{3.2}) is first integrated
with respect to $dr$, which returns the factor
$\mu u^\alpha u^\beta / r^2 u^r$ multiplying 
$\delta[t'-t(r)]\delta(\cos\theta'-1)
\delta(\phi')$. Contractions with $n_\alpha$ are then
taken and the result is substituted into Eq.~(\ref{3.1}).
After simplification, the result is
\begin{equation}
p^2(r) T(r) = G \hat{g}(r) e^{i\omega t(r)},
\label{3.5}
\end{equation}
where
\begin{equation}
G = -\frac{\mu}{8M^2} \biggl[ 
\frac{(\ell-1)\ell(\ell+1)(\ell+2)(2\ell+1)}{4\pi}
\biggr]^{1/2}
\label{3.6}
\end{equation}
if $m=0$, and $G=0$ otherwise ($m$ is the spherical-harmonic
index; that only modes with $m=0$ contribute to the full
perturbation reflects the axial symmetry of the problem). 
Also,
\begin{equation}
\hat{g}(r) = \frac{1}{x(x+1)^2}.
\label{3.7}
\end{equation}

The general solution to the inhomogeneous Teukolsky
equation is obtained by substituting Eq.~(\ref{3.5})
into (\ref{1.3}). The result is
\begin{eqnarray}
R(r) &=& \frac{G}{W_{\rm T}} \Biggl\{ R^\infty(r) 
\biggl[
A + \int_a^r \hat{g}(r') R^H(r') e^{i\omega t(r')}\, dr'
\biggr] 
\nonumber \\ & & \mbox{}
+ R^H(r) \biggl[B + \int_r^b \hat{g}(r') 
R^\infty(r') e^{i\omega t(r')}\, dr' \biggr] \Biggr\}.
\label{3.8}
\end{eqnarray}
Here, $W_{\rm T} \equiv W_{\rm T}(R^H,R^\infty) = 
-2 i \omega^3 {\cal Q}^{\rm in}$, and $a$, $b$,
$A$, and $B$ are constants. 

Our task now is to see to it that the boundary conditions 
--- waves ingoing at the horizon and outgoing at infinity 
--- are properly imposed. In fact, there is no difficulty 
in demanding the correct behavior at the black-hole horizon. 
A short calculation indeed reveals that when $a=2M$ and $A=0$, 
the first term of Eq.~(\ref{3.8}) is $O(f^3)$. Since the second
integral is finite, this ensures that $R(r) \propto R^H(r)$ 
when $r \to 2M$, as required. Unfortunately, the behavior 
at infinity cannot so easily be controlled. This is because 
both integrals diverge when $r\to\infty$, due to the fact 
that $\hat{g}(r) = O(r^{-3/2})$ while $R^{H,\infty}(r) 
= O(r^3)$. Clearly, the solution (\ref{3.8}) must be 
regularized before an attempt is made to impose the 
correct boundary condition at infinity. 

\subsection{Regularization}

I begin by defining the integrals
\begin{equation}
I^A(1,2) = \int_1^2 \hat{g}(r) R^A(r) e^{i\omega t(r)}\, dr,
\label{3.9}
\end{equation}
where the index $A$ stands for either ``$H$'' or 
``$\infty$''. For the purpose of regularization, 
$R^A(r)$ is conveniently expressed as 
\begin{equation}
R^A(r) = \chi^A\, {\cal C} X^A(r),
\label{3.10}
\end{equation}
where $\cal C$ is given in Eq.~(\ref{2.37}). 
Substitution gives 
\begin{equation}
I^A(1,2) = \chi^A (I_{\rm conv} + I_{\rm div}),
\label{3.11}
\end{equation}
where
\begin{eqnarray}
I_{\rm conv} &=& \int_1^2 \Gamma_{\rm conv}(r) 
e^{i\omega t(r)} X^A(r)\, dr,
\label{3.12} \\
I_{\rm div} &=& \int_1^2 \Gamma_{\rm div}(r) 
e^{i\omega t(r)} {\cal L} X^A(r)\, dr,
\label{3.13}
\end{eqnarray}
and
\begin{eqnarray}
\Gamma_{\rm conv}(r) &=& (i\omega r) f \bigl[ \ell(\ell+1) 
- 6M/r\bigr] \hat{g}(r),
\label{3.14} \\
\Gamma_{\rm div}(r) &=& 2 (i\omega r) 
(1-3M/r+i\omega r) r \hat{g}(r).
\label{3.15}
\end{eqnarray}
As the names indicate, $I_{\rm conv}$ is convergent when
$r \to \infty$, since $\Gamma_{\rm conv}(r) = O(r^{-1/2})$, 
while $I_{\rm div}$ is divergent, since 
$\Gamma_{\rm div}(r) = O(r^{3/2})$.

Regularization of $I_{\rm div}$ can be achieved by integration
by parts. To identify what must be done, consider the alternative
form
\begin{eqnarray}
I_{\rm div} &=& \int_1^2 \biggl[ \Gamma_{\rm div} 
e^{i\omega t} {\cal L} X^A + \frac{d}{dr} 
\Bigl( h e^{i\omega t} {\cal L} X^A \Bigr) \biggr]\, dr
- h e^{i\omega t} {\cal L} X^A \biggr|^2_1
\nonumber \\
&\equiv& I'_{\rm div} + \mbox{boundary terms},
\label{3.16}
\end{eqnarray}
where $h(r)$ is a function to be determined. After 
simplification, the new integral becomes 
\begin{equation}
I'_{\rm div} = \int_1^2 e^{i\omega t} 
\bigl( \Gamma'_{\rm div} {\cal L} X^A
+ \Gamma'_{\rm conv} X^A \bigr)\, dr,
\label{3.17}
\end{equation}
where
\begin{equation}
\Gamma'_{\rm div} = \frac{dh}{dr} + i\omega
\biggl( \frac{dt}{dr} + \frac{1}{f} \biggr) h
+ \Gamma_{\rm div},
\label{3.18}
\end{equation}
and 
\begin{equation}
\Gamma'_{\rm conv} = 
\bigl[\ell(\ell+1)-6M/r\bigr]\, \frac{h}{r^2}.
\label{3.19}
\end{equation}
The function $h(r)$ must be chosen so that $I'_{\rm div}$ 
is well behaved when $r\to\infty$.

The divergence of $I_{\rm div}$ is caused by the bad 
behavior of $\Gamma_{\rm div}(r)$. Happily, its 
contribution to $I'_{\rm div}$ can be 
removed by simply setting $\Gamma'_{\rm div} = 0$, 
which gives a differential equation for $h(r)$. One
solution to this equation is
\begin{equation}
h(r) = 8M^2\, \frac{1+x+2Qx^3}{1+x},
\label{3.20}
\end{equation}
where $x=(r/2M)^{1/2}$. Substituting this into 
Eq.~(\ref{3.19}) then reveals that 
$\Gamma'_{\rm conv}(r) = O(r^{-1})$, 
which implies that $I'_{\rm div}$ is 
indeed well behaved. 

Regularization has thus been achieved. Combining 
Eqs.~(\ref{3.11}), (\ref{3.12}), (\ref{3.16}), 
(\ref{3.17}), and (\ref{3.19}) gives
\begin{eqnarray}
I^A(1,2) &=& \chi^A \int_1^2 g(r) e^{i\omega t(r)} X^A(r)\, dr
- \chi^A h(r) e^{i\omega t(r)} {\cal L} X^A(r) \biggr|^2_1
\nonumber \\
&\equiv& \chi^A \Bigl[J^A(1,2) + \mbox{boundary terms}\Bigr],
\label{3.21}
\end{eqnarray}
where
\begin{equation} 
g(r) = \frac{2(1+Qx^3)}{x^4}\, \bigl[ \ell(\ell+1) 
- 6M/r \bigr].
\label{3.22}
\end{equation}
This result will now be put to use.

With the notation introduced above, Eq.~(\ref{3.8}) reads
\begin{equation}
R(r) = \frac{G}{W_{\rm T}} \Bigl\{ R^\infty(r) \bigl[
A + I^H(a,r) \bigr] 
+ R^H(r) \bigl[ B + I^\infty(r,b) \bigr] \Bigr\}.
\label{3.23}
\end{equation}
The regularized version of this equation is obtained by
substituting Eq.~(\ref{3.21}). The boundary terms at
$r'=a$ and $r'=b$ can be absorbed into the constants
$A$ and $B$ by making the replacements
\begin{eqnarray}
A &\to& A + \chi^H h(a) e^{i\omega t(a)} {\cal L} X^H(a),
\nonumber \\
& & \label{3.24} \\
B &\to& B - \chi^\infty h(b) e^{i\omega t(b)} 
{\cal L} X^\infty(b). \nonumber
\end{eqnarray}
The boundary terms at $r'=r$ combine to give
\begin{equation}
J(r) = h e^{i\omega t} \bigl( 
\chi^\infty R^H {\cal L} X^\infty -
\chi^H R^\infty {\cal L} X^H \bigr).
\label{3.25}
\end{equation}
This can be simplified by expressing $R^{H,\infty}(r)$ 
in terms of $X^{H,\infty}(r)$, as in Eq.~(\ref{3.10}). 
After simplification, Eq.~(\ref{3.25}) becomes
\begin{equation}
J(r) = \chi^H \chi^\infty W_{\rm RW} (i\omega r) f
\bigl[\ell(\ell+1)-6M/r \bigr] h(r) e^{i\omega t(r)},
\label{3.26}
\end{equation}
where $W_{\rm RW} \equiv W_{\rm RW}(X^H,X^\infty) =
2 i \omega {\cal A}^{\rm in}$.

Finally, gathering the results yields
\begin{equation}
R(r) = \frac{G}{W_{\rm T}} \biggl\{ R^\infty(r) \Bigl[
A + \chi^H J^H(a,r) \Bigr] 
+ R^H(r) \Bigl[
B + \chi^\infty J^\infty(r,b) \Bigr] + J(r) \biggr\},
\label{3.27}
\end{equation}
where $J(r)$ is given by Eq.~(\ref{3.26}), and
\begin{equation}
J^A(1,2) = \int_1^2 g(r') e^{i\omega t(r')} 
X^A(r')\, dr',
\label{3.28}
\end{equation}
as was first written in Eq.~(\ref{3.21}). 

The function $R(r)$ is now expressed in terms of integrals 
that are well behaved when $r\to\infty$. There is therefore 
no obstacle in setting
\begin{equation}
a = 2M, \qquad b=\infty,
\label{3.29}
\end{equation}
which I shall do from now on. It can then be verified that with 
this choice, the replacements of Eq.~(\ref{3.24}) 
take the form $A\to A+0$, and $B \to B + \infty$. The infinite
shift in $B$ reflects the fact that the original expression 
for $R(r)$, given by Eq.~(\ref{1.2}), was not well defined. 
This shows the importance of the procedure carried out here:
the integrals must be regularized {\it before} $b$ is set
equal to infinity. The fact that $B$ is then shifted by an
infinite amount is of no consequence: Since Eq.~(\ref{3.27})
is a general solution to the inhomogeneous Teukolsky equation, 
as can be verified by direct substitution, this equation is
a perfectly valid starting point for the discussion of boundary
conditions, to which I turn next. One might just as well
forget how Eq.~(\ref{3.27}) was derived, and proceed afresh from
here. 

\subsection{Behavior near $r=2M$}

The behavior of $R(r)$, as expressed by Eq.~(\ref{3.27}), must
now be examined near $r=2M$, to ensure that it correctly represents
purely ingoing waves at the black-hole horizon. The constants
$A$ and $B$ must therefore be chosen so that $R(r\to\infty)
\sim (\mbox{constant})f^2 \exp(-i\omega r^*)$.

Our first task is to evaluate $J^H(2M,r)$ in 
the limit $r\to 2M$. Because $R^\infty(r) = O(f^0)$, 
this calculation must be carried out to second order
in $f$. On the other hand, only the
leading-order term in $J^\infty(r,\infty)$ is required for 
the calculation; this is simply given by $J^\infty(2M,\infty)$. 
Finally, $J(r)$ will have to be computed, also to second
order in $f$.

\subsubsection{Evaluation of $J^H(2M,r)$}

The results of Sec.~II A imply that near $r=2M$, 
$X^H(r)$ can be written as
\begin{equation}
X^H(r) = Y(f) e^{-i\omega r^*},
\label{3.30}
\end{equation}
where $Y(f) = 1 + a_1 f + a_2 f^2 + O(f^3)$. The
coefficients $a_1$ and $a_2$ are given by Eqs.~(\ref{2.7}).
Substituting this into Eq.~(\ref{3.28}) returns an exponential
factor of the form $\exp(i\omega u)$, where $u(r)$ is defined
by
\begin{equation}
u(r) = t(r)-r^* 
= -4M \Bigl[ {\textstyle \frac{1}{3}} x^3
+ {\textstyle \frac{1}{2}} x^2 + x + \ln(x-1) \Bigr],
\label{3.31}
\end{equation}
where $x=(r/2M)^{1/2}$. Changing the integration variable, 
what must be evaluated is
\begin{equation}
J^H(2M,r) = \int_u^\infty \frac{g f Y}{x+1}\, 
e^{i\omega u'}\, du',
\label{3.32}
\end{equation}
where the integrand is considered to be a function of
$u'$. 

To compute the integral, Eq.~(\ref{3.31}) must first be inverted 
in order to express $x$ as a function of $u$. While this cannot be
done exactly in closed form, what is required here is a result
accurate only to second order in
$f=O(x-1)=O(e^{-u/4M})$. Equation (\ref{3.31}) implies
\begin{equation}
U \equiv \exp \biggl(-\frac{u}{4M}- \frac{11}{6} \biggr) 
= (x-1) + 3(x-1)^2 + O\bigl[(x-1)^3\bigr],
\label{3.33}
\end{equation}
which can be inverted to give
\begin{equation}
x-1 = U - 3U^2 + O(U^3).
\label{3.34}
\end{equation}
Next, the integrand is expanded in powers of $x-1$, and 
Eq.~(\ref{3.34}) is used to write this in terms of $U$. 
Integration is then straightforward. The final result 
must be expressed as an expansion in powers of $f$. For 
this purpose one uses $U=\frac{1}{2} f + \frac{9}{8} f^2 
+ O(f^3)$, which follows from $x-1 = \frac{1}{2} f + 
\frac{3}{8} f^2 + O(f^3)$. The final result is
\begin{equation}
J^H(2M,r) = 4M \bigl[ \mu_1 f + \mu_2 f^2 + 
O(f^3) \bigr] e^{i\omega u(r)},
\label{3.35}
\end{equation}
where
\widetext
\begin{eqnarray}
\mu_1 &=& \frac{(1+Q)\bigl[\ell(\ell+1)-3\bigr]}{1-4Q},
\nonumber \\
& & \label{3.36} \\
\mu_2 &=& \frac{2(\ell^4+2\ell^3-5\ell^2-6\ell+12) + 
(2\ell^4+4\ell^3+11\ell^2+9\ell-63)Q + \bigl[ 
6\ell(\ell+1)-42\bigr]Q^2}{4(1-2Q)(1-4Q)}.
\nonumber
\end{eqnarray}
\narrowtext
The evaluation of $J^H(2M,r)$ is now completed.

\subsubsection{Evaluation of $J(r)$}

This calculation is quite straightforward. Equation
(\ref{3.26}), with $h(r)$ given by Eq.~(\ref{3.20}),
can immediately be expanded in powers of $f$. The
result is
\begin{equation}
J(r) = 16M^2 Q \chi^H \chi^\infty W_{\rm RW} 
\bigl[ \nu_1 f + \nu_2 f^2 + O(f^3) \bigr]
e^{i\omega t},
\label{3.37}
\end{equation}
where
\begin{eqnarray}
\nu_1 &=& (1+Q)\bigl[ \ell(\ell+1) - 3 \bigr],
\nonumber \\
& & \label{3.38} \\
\nu_2 &=& \ell(\ell+1) + \frac{3}{4}\,
\Bigl[3\ell(\ell+1)-5\Bigr]Q.
\nonumber
\end{eqnarray}

\subsubsection{Evaluation of $R(r)$}

Expansions in powers of $f$ have been obtained for 
$J^H(2M,r)$ and $J(r)$. These are supplemented by
the results of Sec.~II B and D, which imply
\begin{equation}
R^\infty(r) = {\cal P}^{\rm in} \bigl[ f^2 + 
O(f^3) \bigl] e^{-i\omega r^*} 
+ {\cal P}^{\rm out} \bigl[1 + p^{\rm out}_1 f +
O(f^2) \bigl] e^{i\omega r^*},
\label{3.39}
\end{equation}
where $p^{\rm out}_1$ is given by Eq.~(\ref{2.25}).
We also have $R^H(r) = [f^2 + O(f^3)] \exp(-i\omega
r^*)$ and $J^\infty(r,\infty) = 
J^\infty(2M,\infty) + O(f)$. 

Substituting all this into Eq.~(\ref{3.27}) gives
$R(r)$ as an expansion in powers of $f$, which contains 
terms of order $f^0$, $f$, as well as the allowed terms 
of order $f^2$ and higher. Each term will be discussed 
in turn.

The term of order $f^0$ can be eliminated by setting
\begin{equation}
A =0,
\label{3.40}
\end{equation}
which will be done from here on. 

The term of order $f$ can be simplified using 
Eqs.~(\ref{2.44}), (\ref{2.48}), (\ref{2.53}), 
(\ref{2.55}), (\ref{3.36}), and (\ref{3.38}). 
As it must, it vanishes identically.

The term of order $f^2$ survives, and contains
two contributions. The first is proportional to 
$\exp(i\omega t)$ and comes from $J^H(2M,r)$ and 
$J(r)$; the other is proportional to 
$\exp(-i\omega r^*)$ and comes from $R^H(r)$. 
The first contribution is simplified using the same 
equations as before, in addition to Eq.~(\ref{2.25}); 
the result is $4MQ(1+Q){\cal A}^{\rm in} f^2 e^{i\omega t}$.
This is then combined with the second contribution by
observing that $t=-r^* - 2M(\frac{5}{3}-2\ln 2) +
O(f)$.

After simplification, the final result is that
near $r=2M$,
\begin{equation}
R(r) = \frac{G}{W_{\rm T}} \Bigl[B + C + 
\chi^\infty J^\infty(2M,\infty) \Bigr]\, 
f^2 e^{-i\omega r^*} 
+ O(f^3),
\label{3.41}
\end{equation}
as required. It is recalled that $J^\infty(2M,\infty)$ 
was defined in Eq.~(\ref{3.28}), and the constant $C$ 
is given explicitly by
\begin{equation}
C = 4MQ(1+Q) {\cal A}^{\rm in} \exp\biggl[
-2Q \biggl( \frac{5}{3} - 2\ln 2 \biggr) \biggr].
\label{3.42}
\end{equation}
The constant $B$ will shortly be set to zero.

\subsection{Behavior near $r=\infty$}

It is much easier to extract the behavior of 
Eq.~(\ref{3.27}) near $r=\infty$. I begin with the
computation of $J^\infty(r,\infty)$. In this limit, 
Eq.~(\ref{3.22}) reduces to
\begin{equation}
g(r) = \frac{2Q\ell(\ell+1)}{x}\, \Bigl[1 + O(x^{-2}) \Bigr],
\label{3.43}
\end{equation}
while Eqs.~(\ref{2.9}) and (\ref{2.11}) imply
\begin{equation}
X^\infty(r) = \bigl[1 + O(x^{-2}) \bigr]\, e^{i\omega r^*}.
\label{3.44}
\end{equation}
After substitution into Eq.~(\ref{3.28}), and
a change of integration variable to
\begin{equation}
v(r) = t(r)+r^* 
= -4M \bigl[ {\textstyle \frac{1}{3}} x^3
- {\textstyle \frac{1}{2}} x^2 + x - \ln(x+1) \bigr],
\label{3.45}
\end{equation}
one arrives at
\begin{eqnarray}
J^\infty(r,\infty) &=& 2\ell(\ell+1) Q \int_{-\infty}^v
\frac{1}{x^2} \Bigl[1+O(x^{-1})\Bigr]\, e^{i\omega v'}\, dv' 
\nonumber \\
&=& 4\ell(\ell+1)Q^2 z \bigl[1 + O\bigl(z^{1/2}\bigr) \bigr]\,
e^{i\omega v},
\label{3.46}
\end{eqnarray}
where $z=(i\omega r)^{-1}$.

The computation of $J(r)$ is also straightforward.
Equations (\ref{3.20}) and (\ref{3.26}) immediately give 
\begin{equation}
J(r) = 8M^2 \chi^H \chi^\infty W_{\rm RW} \ell(\ell+1)\,
\frac{1}{z^2}\, \Bigl[1 + O(z) \Bigr]\, e^{i\omega t}.
\label{3.47}
\end{equation}

To finish the job, Eqs.~(\ref{3.46}) and (\ref{3.47}), 
together with the relations $J^H(2M,r) = J^H(2M,\infty) 
+ O(z^{1/2})$ and $R^\infty(r) = [z^{-3}+O(z^{-2})] 
\exp(i\omega r^*)$, are substituted into Eq.~(\ref{3.27}). 
This gives
\begin{equation}
R(r) = \frac{G}{W_{\rm T}} \Bigl\{ 
\chi^H J^H(2M,\infty) \bigl[1 + O\bigl(z^{1/2}\bigr)\bigr] 
R^\infty(r) + B R^H(r) + \mbox{other terms} \Bigr\},
\label{3.48}
\end{equation}
where the ``other terms'' are all proportional to 
$\exp[i\omega t(r)]$, and are $O(z^{-2})$ or higher, and 
therefore much smaller than the dominant terms of order 
$z^{-3}$. Furthermore, because $t(r) \sim 
-(4M/3)(r/2M)^{3/2}$ when $r\to\infty$, their phase 
increases much more rapidly than $r^* \sim r$, which 
means that the ``other terms'' cannot be combined into 
a term proportional to $R^H(r)$, which would represent 
a free, initially incoming, gravitational wave. The only 
such term present in $R(r)$ is $B R^H(r)$, and the 
requirement that waves must be purely outgoing at 
infinity dictates
\begin{equation}
B=0.
\label{3.49}
\end{equation}

The final result is that near $r=\infty$,
\begin{equation}
R(r) = \frac{G}{W_{\rm T}}\, \chi^H
J^H(2M,\infty)\, (i\omega r)^3 e^{i\omega r^*} 
+ O(r^{5/2}),
\label{3.50}
\end{equation}
as required; $J^H(2M,\infty)$ was defined in 
Eq.~(\ref{3.28}).

\section{Summary and conclusion} 

I now summarize. The regularization of the Teukolsky 
equation was successful. The procedure consisted of two 
stages. In the first stage, the most general solution 
to the inhomogeneous Teukolsky equation was written in 
terms of integrals that are well behaved when $r\to\infty$. 
This was accomplished in Eq.~(\ref{3.27}).  In the second 
stage, an ingoing-wave boundary condition was imposed at the
black-hole horizon, and an outgoing-wave boundary
condition was imposed at infinity. While the correct
behavior at infinity could not be verified with the
original form of the solution given by Eq.~(\ref{3.8}), 
it was quite straightforward to do so with the regularized 
form (\ref{3.27}).

The regularized solution is written as
\begin{equation}
R(r) = \frac{G}{W_{\rm T}} \Bigl[ \chi^H J^H(r) R^\infty(r) 
+ \chi^\infty J^\infty(r) R^H(r) + J(r) \Bigr],
\label{4.1}
\end{equation}
where 
\begin{eqnarray}
G &=& -\frac{\mu}{8M^2} \sqrt{\frac{2\ell+1}{4\pi}}
%\nonumber \\ & & \mbox{} \times
\sqrt{(\ell-1)\ell(\ell+1)(\ell+2)},
\label{4.2} \\
W_{\rm T} &=& -2 i \omega^3\, \frac{(1-2Q)(1-4Q)}{4Q}\,
{\cal A}^{\rm in},
\label{4.3} \\
\chi^H &=& \frac{(1-2Q)(1-4Q)}{Q\bigl[ 
(\ell-1)\ell(\ell+1)(\ell+2)-12Q \bigr]}, 
\label{4.4} \\
\chi^\infty &=& \frac{1}{4},
\label{4.5} \\
J^H(r) &=& \int_{2M}^r g(r') e^{i\omega t(r')} 
            X^H(r')\, dr',
\label{4.6} \\
J^\infty(r) &=& \int_r^\infty g(r') e^{i\omega t(r')} 
                 X^\infty(r')\, dr',
\label{4.7} \\
J(r) &=& -2\omega^2 \chi^H \chi^\infty {\cal A}^{\rm in}\,
rf \bigl[ \ell(\ell+1)-6M/r \bigr] h(r) e^{i\omega t(r)}.
\label{4.8}
\end{eqnarray}
The functions $X^H(r)$ and $X^\infty(r)$ are linearly
independent solutions of the Regge-Wheeler equation
(\ref{2.1}), with normalizations determined by 
Eqs.~(\ref{2.42}) and (\ref{2.43}). The constant
${\cal A}^{\rm in}$ is also defined by these equations.
Also, $f=1-2M/r$, $x=(r/2M)^{1/2}$, $Q=iM\omega$, and
\begin{eqnarray}
g(r) &=& \frac{2(1+Qx^3)}{x^4}\, 
\bigl[ \ell(\ell+1)-6M/r \bigr],
\label{4.9} \\
h(r) &=& 8M^2\, \frac{1+x+2Qx^3}{1+x},
\label{4.10} \\
t(r) &=& -2M \biggl( {\textstyle \frac{2}{3}} x^3 + 2x +
\ln \frac{x-1}{x+1} \biggr).
\label{4.11}
\end{eqnarray}

Equation (\ref{4.1}) is obtained directly from 
(\ref{3.27}) by setting $a=2M$, $b=\infty$, $A=B=0$,
$J^H(r) \equiv J^H(2M,r)$, and $J^\infty(r) \equiv
J^\infty(r,\infty)$. Equation (\ref{4.2}) is the
same as (\ref{3.6}). Equation (\ref{4.3}) follows
from (\ref{2.49}) and (\ref{2.53}). Equations
(\ref{4.4}) and (\ref{4.5}) are the same as 
(\ref{2.48}). Equations (\ref{4.6}) and (\ref{4.7}) 
are the same as (\ref{3.28}). And finally, 
Eqs.~(\ref{4.8})--(\ref{4.11}) are 
the same as (\ref{3.26}), (\ref{3.22}), (\ref{3.20}),
and (\ref{3.4}), respectively.

Equation (\ref{4.1}) implies that near $r= 2M$,
$R(r)$ behaves as
\begin{equation}
R(r) \sim \frac{G}{W_{\rm T}}\, \Bigl[
2^{2+4Q} e^{-10Q/3} M Q(1+Q) {\cal A}^{\rm in} +
\chi^\infty J^\infty(2M) \Bigr]\, f^2 e^{-i\omega r^*}.
\label{4.12}
\end{equation}
This follows from Eqs.~(\ref{3.41}) and (\ref{3.42}).
It also implies that near $r=\infty$,
\begin{equation}
R(r) \sim \frac{G}{W_{\rm T}}\, \chi^H
J^H(\infty)\, (i\omega r)^3 e^{i\omega r^*},
\label{4.13}
\end{equation}
which is the same statement as in Eq.~(\ref{3.50}). This 
expression agrees precisely with the one derived by Simone, 
Poisson, and Will \cite{SimonePoissonWill}, who obtained it 
by ``throwing away the infinite boundary term''.

The radial function $R(r)$, found here to be a solution of the 
inhomogeneous Teukolsky equation, represents a gravitational 
perturbation of frequency $\omega$ and spherical-harmonic 
indices $\ell$ and $m$. A better notation for it (which I did 
not adopt in order to keep all symbols simple) would be 
$R_{\ell m}(\omega;r)$. The full perturbation is obtained by 
summing over all these modes. More precisely, the perturbation 
in the Riemann tensor caused by the infalling particle is
represented by the complex function $\Psi_4$ 
\cite{NewmanPenrose} given by
\begin{equation}
\Psi_4(x) = \frac{1}{r^4} \int \sum_{\ell m} 
R_{\ell m}(\omega;r)\, \mbox{}_{-2}Y_{\ell m}
(\theta,\phi) e^{-i\omega t}\, d\omega.
\label{4.14}
\end{equation}
Here, $x$ is the event in spacetime labeled by the Schwarzschild 
coordinates $(t,r,\theta,\phi)$, and the functions  
$\mbox{}_{-2}Y_{\ell m}(\theta,\phi)$ are spherical
harmonics of spin-weight $-2$ \cite{Goldbergetal}. For 
an axially symmetric problem, such as the one considered 
in this paper, modes with $m\neq 0$ vanish identically, 
so the sum over $m$ reduces to the single term $m=0$. 
From $\Psi_4(x)$ one may obtain many relevant quantities, 
such as the gravitational-wave field $h^{\rm TT}_{ab}(x)$ 
and the fluxes of energy at the black-hole horizon and at 
infinity. Additional details are provided by 
Refs.~\cite{15,SimonePoissonWill}.

The considerations of this paper were limited to the simplest 
case of an infall into a black hole: the hole was assumed to 
be nonrotating, and the particle was assumed to have zero 
angular momentum and a vanishing initial velocity. There is,
however, no reason to believe that the methods used here 
could not be extended to more complicated situations. Of 
course, the amount of labor involved, already considerable 
here, would increase, but there is no issue of principle. 

To conclude, I would like to stress the main message
of this paper. The standard choice of Green's function
for solving the inhomogeneous Teukolsky equation leads
to divergent integrals, and contrary to naive expectations,
fails to enforce the correct boundary conditions at the 
black-hole horizon and at infinity. The regularization
procedure amounts to nothing more --- and nothing
less --- than finding an adequate Green's function. Contrary
to what may have been believed, there is nothing intrinsically
wrong with the Teukolsky equation when dealing with non-compact
source terms. 

\section*{Acknowledgments}

The work presented in this paper was conceived during a 
conversation with Steve Detweiler, who also was wondering 
why it was necessary to ``throwing away the infinite 
boundary term''. This work was supported by
the Natural Sciences and Engineering Research Council
of Canada, by the National Science Foundation
under Grant No.~PHY 92-22902, and by the
National Aeronautics and Space Administration under
Grant No.~NAGW 3874.


\begin{references}

\bibitem[*]{email} E-mail address: 
        {\tt poisson@terra.physics.uoguelph.ca}.
\bibitem[\dagger]{pa} Permanent address.
\bibitem{DRPP} M. Davis, R. Ruffini, W.H. Press, and R.H. Price,
        Phys. Rev. Lett. {\bf 27}, 1466 (1971); M. Davis and
        R. Ruffini, Lett. Nuovo Cimento {\bf 2}, 1165 (1971);
        M. Davis, R. Ruffini, and J. Tiomno, Phys. Rev. D 
        {\bf 5}, 2932 (1972).
\bibitem{Zerilli} F.J. Zerilli, Phys. Rev. D {\bf 2}, 2141
        (1970).
\bibitem{Ruffini} R. Ruffini, Phys. Rev. D {\bf 7}, 972 (1973);
        R. Ruffini and M. Sasaki, Prog. Theor. Phys. {\bf 66},
        1627 (1981).
\bibitem{DetweilerSzedenits} S.L. Detweiler and E. Szedenits,
        Astrophys. J. {\bf 231}, 211 (1979).
\bibitem{SasakiNakamura82} M. Sasaki and T. Nakamura, Prog.
        Theor. Phys. {\bf 67}, 1788 (1982); T. Nakamura and
        M. Haugan, Astrophys. J. {\bf 269}, 292 (1983).
\bibitem{KojimaNakamura83} Y. Kojima and T. Nakamura,
        Phys. Lett. {\bf 96A}, 335 (1983).
\bibitem{KojimaNakamura84} Y. Kojima and T. Nakamura,
        Prog. Theor. Phys. {\bf 71}, 79 (1984).
\bibitem{MinoShibataTanaka} Y. Mino, M. Shibata, and
        T. Tanaka, Phys. Rev. D {\bf 53}, 622 (1996).
\bibitem{1} E. Poisson, Phys. Rev. D {\bf 47}, 1497 (1993).
\bibitem{2} C. Cutler, L.S. Finn, E. Poisson, and 
        G.J. Sussman, Phys. Rev. D {\bf 47}, 1511 (1993).
\bibitem{3} T. Tanaka, M. Shibata, M. Sasaki, H. Tagashi, 
        and T. Nakamura, Prog. Theor. Phys. {\bf 90}, 65 
        (1993).
\bibitem{4} T. Apostolatos, D. Kennefick, 
        A. Ori, and E. Poisson, Phys. Rev. D {\bf 47}, 
        5376 (1993).
\bibitem{5} M. Shibata, Phys. Rev. D {\bf 48}, 663
        (1993).
\bibitem{6} E. Poisson, Phys. Rev. D {\bf 48},
        1860 (1993).
\bibitem{7} M. Shibata, Prog. Theor. Phys. {\bf 90}, 
        595 (1993).
\bibitem{8} H. Tagoshi and T. Nakamura,
        Phys. Rev. D {\bf 49} 4016 (1994).
\bibitem{9} C. Cutler, D. Kennefick, and E. Poisson,
        Phys. Rev. D {\bf 50}, 3816 (1994).
\bibitem{10} M. Sasaki, Prog. Theor. Phys. {\bf 92}, 17 
        (1994).
\bibitem{11} H. Tagoshi and M. Sasaki, Prog.
        Theor. Phys. {\bf 92}, 745 (1994).
\bibitem{12} M. Shibata, Phys. Rev. D {\bf 50}, 6297 (1994).
\bibitem{13} M. Shibata, M. Sasaki, H. Tagoshi,
        and T. Tanaka, Phys. Rev. D {\bf 51}, 1646 (1995).
\bibitem{14} H. Tagoshi, Prog. Theor. Phys. {\bf 93},
         307 (1995).
\bibitem{15} E. Poisson and M. Sasaki, Phys. Rev. D
        {\bf 51}, 5753 (1995).
\bibitem{16} E. Poisson, Phys. Rev. D {\bf 52}, 5719
        (1995). 
\bibitem{17} H. Tagoshi, M. Shibata, T. Tanaka, and M.
        Sasaki, {\it Post-Newtonian expansion of 
        gravitational waves from a particle in circular
        orbit around a rotating black hole: Up to $O(v^8)$
        beyond the quadrupole formula}, Phys. Rev. D
        (to be published, July 15 1996).
\bibitem{18} T. Tanaka, Y. Mino, M. Sasaki, and M. Shibata,
        {\it Gravitational waves from a spinning particle
        in circular orbit around a rotating black hole},
        Phys. Rev. D (to be published, August 15 1996).
\bibitem{Detweiler} S.L. Detweiler, Astrophys. J.
        {\bf 225}, 687 (1978).
\bibitem{Teukolsky} S.A. Teukolsky, Astrophys. J.
        {\bf 185}, 635 (1973).
\bibitem{BardeenPress} J.M. Bardeen and W.H. Press,
        J. Math. Phys. {\bf 14}, 7 (1973).
\bibitem{SimonePoissonWill} L.E. Simone, E. Poisson,
        and C.M. Will, Phys. Rev. D {\bf 52}, 4481
        (1995).
\bibitem{TashiroEzawa} Y. Tashiro and H. Ezawa,
        Prog. Theor. Phys. {\bf 66}, 1612 (1981).
\bibitem{SasakiNakamura81} M. Sasaki and T. Nakamura,
        Phys. Lett. {\bf 87A}, 85 (1981). 
\bibitem{ReggeWheeler} T. Regge and J.A. Wheeler,
        Phys. Rev. {\bf 108}, 1063 (1957).
\bibitem{Arfken} See, for example, G. Arfken, {\it
        Mathematical Methods for Physicists} (Academic
        Press, Orlando, 1985), Section 16.5.
\bibitem{Chandrasekhar} S. Chandrasekhar, Proc. R. Soc.
        London {\bf A343}, 289 (1975).
\bibitem{NewmanPenrose} E.T. Newman and R. Penrose, 
        J. Math. Phys. {\bf 7}, 863 (1966). 
\bibitem{Goldbergetal} J.N. Goldberg, A.J. MacFarlane,
         E.T. Newman, F. Rohrlich and E.C.G. Sudarshan,
         J. Math. Phys. {\bf 8}, 2155 (1967). 
\end{references}
\end{document}